\documentclass[prl,twocolumn,showpacs,preprintnumbers]{revtex4}
\usepackage{textcomp}
\usepackage{amsmath}
\usepackage{amssymb}
\usepackage{graphicx}

\begin{document}

\title{Cold Atom Physics Using Ultra-Thin Optical Fibers:
 Light-Induced Dipole Forces and Surface Interactions}
\author{G.~Sagu\'{e}, E.~Vetsch, W.~Alt, D. Meschede}

\affiliation{Institut f\"ur Angewandte Physik, Universit\"at Bonn,
Wegelerstr.~8, 53115 Bonn, Germany}

\author{A. Rauschenbeutel}

\altaffiliation{Present address: Institut f\"ur Physik,
Universit\"at Mainz, 55099 Mainz, Germany}

\email{rauschenbeutel@uni-mainz.de}

\affiliation{Institut f\"ur Angewandte Physik, Universit\"at Bonn,
Wegelerstr.~8, 53115 Bonn, Germany}

\date{\today}

\begin{abstract}

The strong evanescent field around ultra-thin unclad optical fibers
bears a high potential for detecting, trapping, and manipulating
cold atoms. Introducing such a fiber into a cold atom cloud, we
investigate the interaction of a small number of cold Caesium atoms
with the guided fiber mode and with the fiber surface. Using high
resolution spectroscopy, we observe and analyze light-induced
dipole forces, van der Waals interaction, and a significant
enhancement of the spontaneous emission rate of the atoms. The
latter can be assigned to the modification of the vacuum modes by
the fiber.

\end{abstract}

\pacs{42.50.-p, 39.25.+k, 34.50.Dy, 32.80.Pj}

\maketitle

Tapered and microstructured optical fibers count among the most
active fields of research in recent years \cite{Russell03,Tong03}.
In such fibers, the propagation of light can, e.g., be tailored
such that it controllably depends on the light intensity. These
fiber-induced non-linear processes play for instance a major role
in the generation of optical frequency combs and often stem from
the non-linear response of the bulk fiber material, subjected to
extreme intensities. The low intensity limit of non-linear
light-matter interaction is reached when single photons already
induce a non-linear response of matter. This situation is realized
in cavity quantum electrodynamics \cite{Berman}, where photons,
typically confined in space by an optical resonator \cite{Vahala},
interact with a single or a few dipole emitters.

In this context, ultra-thin unclad optical fibers offer a strong
transverse confinement of the guided fiber mode while exhibiting a
pronounced evanescent field surrounding the fiber \cite{Balykin1}.
This unique combination allows to efficiently couple particles
(atoms, molecules, quantum dots etc.) on or near the fiber surface
to the guided fiber mode, making tapered optical fibers (TOFs) a
powerful tool for their detection, investigation, and
manipulation: The absorbance of organic dye molecules, deposited
on a subwavelength-diameter TOF, has been spectroscopically
characterized via the fiber transmission with unprecedented
sensitivity \cite{Warken07}. Furthermore, the fluorescence from a
very small number of resonantly irradiated atoms around a 400-nm
diameter TOF, coupled into the guided fiber mode, has been
detected and spectrally analyzed \cite{LeKien05,Nayak06}. Finally,
it has also been proposed to trap atoms around ultra-thin fibers
using the optical dipole force exerted by the evanescent field
\cite{Dowling96,Balykin3}.

Here, we report on the observation of such dipole forces induced
by the evanescent field around a sub-wavelength diameter TOF. By
spectroscopically investigating the transmission of a probe laser
launched through the fiber, we find clear evidence of the
mechanical effects of these dipole forces on the atoms, leading to
a modification of the atomic density in the vicinity of the fiber.
A rigorous analysis furthermore shows that a detailed description
of the absorption signal must include the interaction of the atoms
with the dielectric fiber. In particular, this includes the
mechanical and spectral effects of the van der Waals (vdW)
interaction and a significant enhancement of the spontaneous
emission rate of the atoms due to a modification of the vacuum
modes by the fiber. An enhanced spontaneous emission of atoms in
vacuo coupled to a continuous set of evanescent modes has already
been observed in evanescent wave spectroscopy on a plane
dielectric surface \cite{Ivanov04}. In our case, however, the
ultra-thin fiber only sustains a single mode at the atomic
wavelength. To our knowledge, the enhanced spontaneous emission of
atoms in vacuo in such a situation has never been observed before.

\begin{figure}
 \centering
 \includegraphics[width=0.4\textwidth]{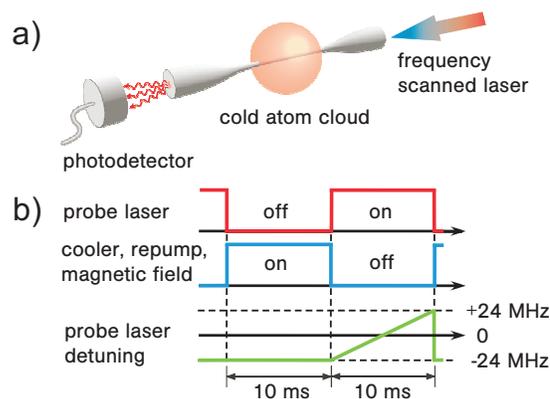}
  \caption{\label{fig:setup} (a) Schematic experimental setup.
  A cloud of laser cooled Caesium atoms is spatially overlapped
  with the 500-nm diameter waist
  of a tapered optical fiber. The transmission through the fiber is
  measured using a photodetector. (b) Timing of the experiment}
\end{figure}

Figure~\ref{fig:setup} (a) shows the schematic experimental setup.
A cloud of cold Caesium atoms, released from a magneto-optical
trap (MOT), is spatially overlapped with the waist of a TOF in an
ultra high vacuum (UHV) environment. The MOT is geometrically
aligned by means of a bias magnetic field while monitoring its
position with two CCD cameras. A frequency scanned probe laser is
launched through the fiber and its transmission is measured with
an avalanche photodiode (APD). Typical powers of the probe laser
used in the experiment range from several hundred Femtowatts to
one Nanowatt.

We fabricate the tapered fibers by stretching a standard single
mode fiber (Newport F-SF) while heating it with a travelling
hydrogen/oxygen flame \cite{Birks92}. Our computer controlled fiber
pulling rig produces tapered fibers with a homogeneous waist
diameter down to 100~nm and a typical extension of 1--10~mm. In the
taper sections, the weakly guided LP$_{01}$ mode of the unstretched
fiber is adiabatically transformed into the strongly guided
HE$_{11}$ mode of the ultrathin section and back \cite{Love86},
resulting in a highly efficient coupling of light into and out of
the taper waist. For TOFs with final diameter above 0.5~$\mu$m, we
achieve up to 97~\% of the initial transmission at 852~nm. For the
present experiment we used a 500-nm diameter fiber with 93\%
transmission and a waist length of 5~mm, sustaining only the
fundamental HE$_{11}$ mode at the 852-nm Cs D2 wavelength. During
evacuation of the vacuum chamber, the fiber transmission dropped to
40~\%, possibly due to contamination with pump oil. After one day
in UHV, the transmission increased again to 80\%.

We use a conventional Cs MOT with a $1/\sqrt{e}$-radius of 0.6~mm.
We probe the atoms with a diode laser which is frequency scanned
by $\pm 24$~MHz with respect to the $6^2S_{1/2}$, $F=4$ to
$6^2P_{3/2}$, $F=5$ transition using an acusto-optical modulator
in double pass configuration. While being linearly polarized
before coupling into the TOF, the probe laser polarization at the
position of the fiber waist is unknown. The probe laser linewidth
of 1~MHz allows to resolve the 5.2-MHz natural linewidth of the Cs
D2 line in Doppler-free spectroscopy.

Figure~\ref{fig:setup} (b) shows the timing of the experimental
sequence: During the first 10~ms, the atoms are captured and
cooled in the MOT while the probe laser is off. In the following
10~ms, the MOT cooling- and repump-laser and the magnetic field
are off and the probe laser is on. The atoms are thus not
influenced by the MOT beams or magnetic fields during the
spectroscopy.
\begin{figure}
 \centering
 \includegraphics[width=0.3\textwidth]{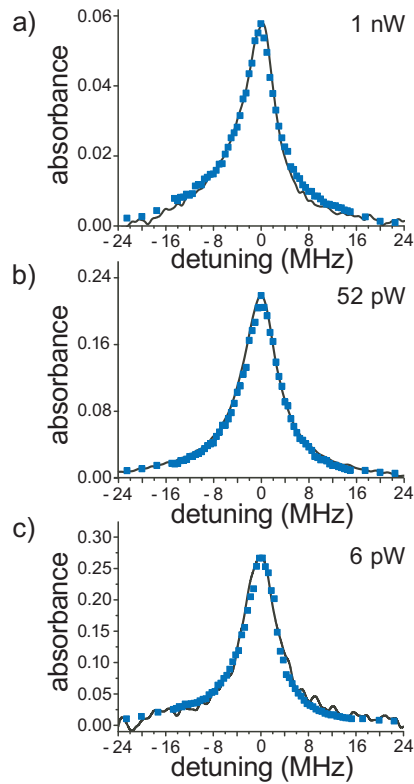}
  \caption{\label{fig:spectra} Measured (line graphs) and simulated
  (squares) absorbance of the atoms
  versus the detuning of the probe
  laser. The effective number of atoms contributing to the spectra
  (see Eq.~(\ref{eq:N})) is 107, 14 and 2 for (a) 1 nW, (b) 52 pW and (c) 6 pW of probe
  laser power, respectively. The decrease of the peak absorbance with increasing
  power is due to the saturation of the atoms.}
\end{figure}
The APD signal is recorded with a digital storage oscilloscope and
averaged over 4096 traces. Figure~\ref{fig:spectra} shows the
measured (line graphs) and theoretically predicted (squares)
absorbance of the atoms (negative natural logarithm of the
transmission) versus the probe laser detuning for three different
probe laser powers. The theory assumes an averaged atomic density
distribution $\rho_{\delta,P}(r,z)$ around the fiber, where $r$ is
the distance from the fiber center and $z$ the position along the
fiber waist. Note that, due to light-induced dipole forces,
$\rho_{\delta,P}$ also depends on the detuning of the probe laser
with respect to resonance, $\delta$, and on its power, $P$. The
line shape is then given by
\begin{equation}\label{eq:A_von_delta}
A_P(\delta)=\frac{\hbar w}{P}\int\rho_{\delta,P}(r,z)\Gamma
(I_{P}(r),\gamma(r),\delta+\delta_\mathrm{vdW}(r))\,dV\, ,
\end{equation}
where $\Gamma (I_{P}(r),\gamma(r),\delta+\delta_\mathrm{vdW}(r))$
is the scattering rate of an atom in the evanescent field with
intensity $I_{P}(r)$, $\gamma(r)$ is the longitudinal decay rate of
the atom, and $\delta_\mathrm{vdW}(r)$ is the vdW shift of the
atomic transition frequency.

The evanescent field intensity profile, $I_{P}(r)$, can be found in
\cite{Balykin1}. The polarization state of the evanescent field has
been assumed to be an incoherent, equally weighted mixture of
linearly and circularly polarized light. Under these conditions,
the Cs saturation intensity in free space is 18~W/m$^2$. The
longitudinal decay rate strongly depends on the atom-fiber
distance~\cite{V. V. Klimov}. Given that the silica fiber is
transparent at the Cs D2 wavelength, $\gamma(r)$ has only two
contributions: emission into freely propagating modes and emission
into guided fiber modes:
\begin{equation}\label{eq:gamma(r)}
\gamma(r)=\gamma_{\mathrm{free}}(r)+\gamma_{\mathrm{guid}}(r)\ .
\end{equation}
For an atom near a 500-nm diameter dielectric cylinder at distances
smaller than the emission wavelength, $\gamma_{\mathrm{free}}(r)$
is given in~\cite{V. V. Klimov} while $\gamma_\mathrm{guid}(r)$ can
be approximated as
\begin{equation}
\gamma_\mathrm{guid}(r)\simeq 0.3\,\gamma_{0}I_P(r)/I_P(a)\ .
\end{equation}
Here, $\gamma_{0}$ is the spontaneous emission rate of a Cs atom in
free space, $a$ denotes the fiber radius, and 0.3\,$\gamma_{0}$
corresponds to the spontaneous emission rate of an atom placed on
the surface of a 500 nm diameter optical fiber into the guided
mode~\cite{LeKien05}. On the surface of the fiber,
Eq.~(\ref{eq:gamma(r)}) then predicts a 57\% increase of the
spontaneous emission rate of the Cs atoms, resulting in a
broadening of the absorbance line shapes.

We calculated the vdW shift, $\delta_\mathrm{vdW}(r)$, for the D2
line of Cs near a 500 nm diameter dielectric
cylinder~\cite{Boustimi02}. It stems from the different
polarizabilities of the $6^2S_{1/2}$ ground state and the excited
$6^2P_{3/2}$ state of the Cs atoms when interacting with the
dielectric surface. According to Eq.~(\ref{eq:A_von_delta}),
$\delta_\mathrm{vdW}(r)$ thus inhomogeneously broadens the
absorbance profile: Atoms at different distances from the fiber
surface will be unequally shifted while contributing to
$A_P(\delta)$. Furthermore, we expect the center of the profile to
be red-shifted by at most $-0.5$~MHz. However, being of the same
order as the drifts of our probe laser frequency, this shift is too
small to be experimentally quantified using the current setup.

Finally, we assume the following explicit form for the density
distribution of the atomic cloud:
\begin{equation}\label{eq:density}
\rho_{\delta,P}(r,z)=\left\{\frac{n_0}{\sigma^3
(2\pi)^{\frac{3}{2}}} e^{-\frac{r^2+z^2}{2\sigma^2}}\right\}
f_{\delta,P}(r)\ .
\end{equation}
Here, the term in curly brackets corresponds to a Gaussian density
distribution of the unperturbed atomic cloud with $\sigma=0.6$~mm
radius, containing $n_0$ atoms. The factor $f_{\delta,P}(r)$
accounts for the perturbation introduced by the presence of the
fiber. We calculate $f_{\delta,P}(r)$ with a Monte Carlo simulation
of 100,000 trajectories of thermal atoms with a temperature of
125~$\mu$K, i.e., the Cs Doppler temperature. This simulation
includes the attractive vdW force between the fiber surface and the
atoms and the saturating dipole force induced by the probe
laser~\cite{J. E. Bjorkholm}.
\begin{figure}
 \centering
 \includegraphics[width=7cm]{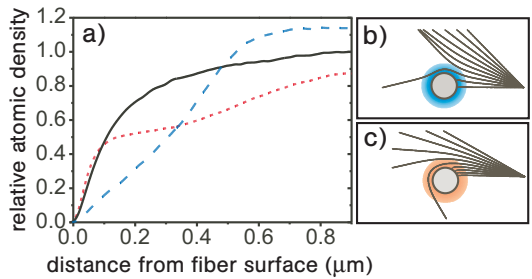}
  \caption{\label{fig:density_simulation}
  (a) Simulations of the relative density of the MOT for different
  detunings $\delta$ of the probe laser versus the distance from the
  fiber surface; solid line $\delta=0$~MHz,
  dotted line $\delta=-3$~MHz, dashed line $\delta=3$~MHz.
  The following parameters have been used
  for the simulations:
  Fiber diameter 500 nm, probe power 1 nW and a 3D Maxwellian velocity distribution of the
  Cs atoms
  at a temperature of 125 $\mathrm{\mu K}$.
  (b) and (c) show several atomic trajectories for $\delta=+3$ MHz and $\delta=-3$ MHz respectively with a
   fixed atom velocity of 10 cm/s.}
\end{figure}
Figure~\ref{fig:density_simulation}(a) shows $f_{\delta,P}(r)$ as a
function of the distance from the fiber surface for $P=1$~nW and
$\delta=-3$, 0, and $+3$~MHz (dotted, solid, and dashed line). The
frequency dependency of $f_{\delta,P}(r)$ due to light-induced
dipole forces is clearly apparent. In all three cases
$f_{\delta,P}(r)$ decays to zero at the surface of the fiber due to
the vdW force.

$A_P(\delta)$ from Eq.~(\ref{eq:A_von_delta}) can now be adjusted
to the experimental line shapes, the only fitting parameters being
$n_0$ and an experimental frequency offset.
Figure~\ref{fig:spectra} shows three examples for $P$ ranging over
three orders of magnitude. The agreement between theory (squares)
and experiment (line graphs) is excellent. In particular, in
addition to the line width, our model reproduces well the asymmetry
of the line shape observed for larger powers.

\begin{figure}
 \centering
 \includegraphics[width=0.35\textwidth]{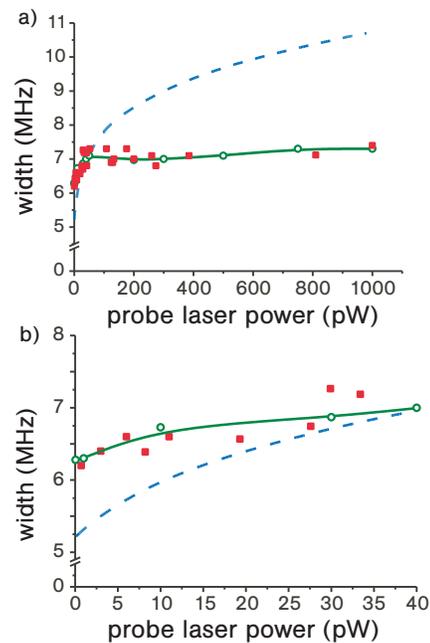}
  \caption{\label{fig:linewidth} Linewidth of the absorbance profiles
  versus probe laser power. (a) full power range
  and (b) low power range. The squares correspond to
  the experimental data and the open circles are the simulated values.
  The continuous line is a guide to the eye (b-spline)
  of the simulated values. The dashed line is the reduced model
  not taking into account light-induced dipole forces and surface
  interactions.}
\end{figure}

Figure~\ref{fig:linewidth} shows the width of the measured
absorbance profiles versus the probe laser power (squares). The
linewidths predicted by our model are also shown (open circles with
a b-spline fit as a guide to the eye). We recall that the effects
of light-induced dipole forces and surface interactions have been
included in the model. For comparison, we also show the expected
linewidths in absence of these effects (dashed line). While the
full model agrees very well with the experimental data, the reduced
model strongly deviates both for high and low powers.

For probe laser powers larger than 100 pW, the measured lines are
considerably narrower than what would be expected in absence of
dipole forces and surface interactions, see
Fig.~\ref{fig:linewidth}(a). For 1 nW of probe laser power this
narrowing exceeds 40\%. The narrowing can be explained by the
effect of the light-induced dipole forces on the density of the
atomic cloud, see Fig.~\ref{fig:density_simulation}(a). For
distances smaller than 370 nm, i.e., in the region that contains
more than 75\% of the evanescent field power, the largest
integrated density of the atomic cloud is predicted in the case of
zero detuning ($\delta=0$~MHz). For blue ($\delta=+3$ MHz) and red
($\delta=-3$ MHz) detunings, this integrated density is lowered due
to the effect of the light-induced dipole forces. This results in a
reduced absorbance and leads to an effective line narrowing.
Figures~\ref{fig:density_simulation}(b) and (c) show several
simulated atomic trajectories with fixed initial velocity. For the
case of blue detuning, (b), the atoms are repelled by the fiber due
to the repulsive light-induced dipole force. For the case of red
detuning, (c), the atoms are accelerated towards the fiber.
Naively, one might assume that this increases the density close to
the fiber. However, this effect is counteracted by the shorter
average time of flight of the atoms through the evanescent field
due to their higher velocity and by the higher atomic loss rate
\cite{loss_rate}. In fact, for distances up to 100~nm both effects
cancel almost perfectly. For larger distances, however, the effects
reducing the density dominate. The net effect is therefore also a
reduction of the absorbance.

Figure~\ref{fig:linewidth}(b) shows the linewidths for the limit
of low probe laser powers, i.e., low saturation and negligible
light-induced dipole forces. The measured linewidths approach
6.2~MHz for vanishing powers. This result exceeds the natural Cs
D2 linewidth in free space by almost 20~\%. This broadening can be
explained by surface interactions, i.e., the vdW shift of the Cs
D2 line and the modification of the spontaneous emission rate of
the atoms near the fiber, see Eq.~(\ref{eq:A_von_delta}). Both
effects have the same magnitude and only their combination yields
the very good agreement between our model and the experimental
data.

Finally, we estimate the effective number, $N_P$, of fully
saturated atoms contributing to the signals in
Fig.~\ref{fig:spectra}. From the adjustment of the height of the
absorbance profiles, we extract the total number of atoms in the
cloud, $n_0$, and infer a maximum atomic density of $4.4\times
10^{10}$ atoms/cm$^3$ using Eq.~(\ref{eq:density}). This value is
slightly smaller than typical peak densities of unperturbed Cs
MOTs~\cite{tow}. We now estimate $N_P$ according to
\begin{equation}\label{eq:N}
N_P=\frac{2}{\gamma_0}\int\rho_{\delta=0,P}(r,z) \Gamma
(I_P(r),\gamma(r),\delta_\mathrm{vdW}(r))\,dV\ ,
\end{equation}
where we follow the notation of Eq.~(\ref{eq:A_von_delta}). Note
that $N_P$ is power dependent and can be lowered by reducing $P$.
We calculate $N_P$ to be 107, 14, and 2 in
Fig.~\ref{fig:spectra}(a), (b), and (c), respectively. Furthermore,
due to the saturating scattering rate $\Gamma$ in the integrand of
Eq.~(\ref{eq:N}), the mean distance of the probed atoms from the
fiber surface is also power dependent and can be adjusted down to
248~nm.

Summarizing, we have shown that sub-wavelength diameter optical
fibers can be used to detect, spectroscopically investigate, and
mechanically manipulate extremely small samples of cold atoms. In
particular, on resonance, as little as two atoms on average,
coupled to the evanescent field surrounding the fiber, already
absorbed 20~\% of the total power transmitted through the fiber.
These results open the route towards the use of ultra-thin fibers
as a powerful tool in quantum optics and cold atom physics. By
optically trapping one or more atoms around such fibers
\cite{Dowling96,Balykin3}, it should become possible to
deterministically couple the atoms to the guided fiber mode and to
even mediate a coupling between two simultaneously trapped atoms
\cite{LeKien05b}, leading to a number of applications, e.g., in the
context of quantum information processing. In addition, high
precision measurements of the modification of the lifetime of
atomic energy levels near surfaces and of the van der Waals
potential \cite{M. Chevrollier} are also within the scope of such
glass fiber quantum optics experiments.

We wish to thank V.~I.~Balykin and D.~Haubrich for their
contribution in the early stages of the experiment, F.~Warken for
assistance in the fiber production, B.~Weise for his part in the
simulations, and M.~Ducloy and C.~Henkel for valuable discussions.
This work was supported by the EC (Research Training Network
``FASTNet'') and the DFG (Research Unit 557).


\begin{thebibliography}{99}
\bibitem{loss_rate} Upon collision with the fiber surface, the atoms are
either adsorbed or rapidly expelled with a kinetic energy of
$E_\mathrm{kin}/k_\mathrm{B}\approx 300$~K and do not contribute to
the absorbance signal anymore.
\end{thebibliography}
\end{document}